\documentclass[10pt, journal]{IEEEtran}
\usepackage[nocompress]{cite}
\usepackage[english]{babel}
\usepackage{url}
\usepackage[labelfont=bf]{caption}

\begin{document}

\title{De-specializing an HLS library for Deep Neural Networks: improvements upon hls4ml}

\author{\IEEEauthorblockN{Serena Curzel,
Nicolò Ghielmetti, Michele Fiorito and
Fabrizio Ferrandi}\\
\IEEEauthorblockA{\textit{Dipartimento di Elettronica, Informazione e Bioingegneria}\\
Politecnico di Milano, Italy\\%
}
}
\maketitle

\begin{abstract}
Custom hardware accelerators for Deep Neural Networks are increasingly popular: in fact, the flexibility and performance offered by FPGAs are well-suited to the computational effort and low latency constraints required by many image recognition and natural language processing tasks.
The gap between high-level Machine Learning frameworks (e.g., Tensorflow, Pytorch) and low-level hardware design in Verilog/VHDL creates a barrier to widespread adoption of FPGAs, which can be overcome with the help of High-Level Synthesis.
hls4ml is a framework that translates Deep Neural Networks into annotated C++ code for High-Level Synthesis, offering a complete and user-friendly design process that has been enthusiastically adopted in physics research.
We analyze the strengths and weaknesses of hls4ml, drafting a plan to enhance its core library of components in order to allow more advanced optimizations, target a wider selection of FPGAs, and support larger Neural Network models.
\end{abstract}

\section{Introduction}

Deep Neural Networks (and other types of Machine Learning models) are data-hungry, computationally intensive algorithms capable of obtaining outstanding results in applications such as image classification and speech recognition.
As new models keep growing in size and complexity, their number of operations and memory footprint pose significant challenges, especially under the strict latency and power consumption constraints typical of real-time edge systems.

Thus, there is a growing need for specialized inference accelerators, with FPGA-based solutions often providing a good trade-off between flexibility and performance.
FPGAs offer a faster design time than ASICs, reconfigurability allows them to keep up with the ongoing evolution of Neural Network models, and they consume significantly less power than GPUs.
They can be used to create either processor-like systems that support many types of Neural Networks or specialized accelerators tailored to a single network model.

The main obstacle to broader adoption of FPGAs in the Deep Learning community is the wide gap between software design and RTL implementation: the translation of a Deep Neural Network model (typically written and trained in a Python-based software framework) into an efficient FPGA design requires hardware expertise that is not easily found among Machine Learning experts.
Automatic design flows have been proposed to mitigate this issue; many exploit well-established High-Level Synthesis techniques to translate annotated C/C++ code into Verilog/VHDL.

One HLS-based framework is hls4ml \cite{duarte2018fast}, which was first developed to enhance high energy physics experiments at CERN by simplifying the development of Deep Learning accelerators on FPGA.
hls4ml is rapidly evolving; however, its core functionality remains tied to a library of C++ components representing standard Deep Learning operators.
This library has been optimized for Vivado HLS (Xilinx's High-Level Synthesis tool), it contains parametric data types to allow quantization, and it assumes that the accelerator will be composed of a sequence of operators representing the different network layers.

We argue that such characteristics limit the real applicability of hls4ml to state-of-the-art Deep Neural Network models and propose modifications to the library of components that have the potential of making it more portable and efficient.
We also aim to substitute commercial tool Vivado HLS with an open-source High-Level Synthesis engine, Bambu \cite{panda}, to have more control on the translation from C++ to Verilog/VHDL and to increase the number of supported FPGA targets.
Part of this work has only been outlined in theory, while part of it is currently under development, and we expect to support our claims with experimental evidence within a short time frame.

\section{HLS tools and compilers for Deep Learning}

Extensive research has been published on specialized processors for Deep Learning, and many of them were implemented on FPGA.
With a fixed processor template, different applications can achieve different performance depending on how well they exploit the available computational resources.
Instead, our focus in this section is on tools and compilers to support the development of accelerators tuned to a specific application and a specific Neural Network model, sacrificing flexibility to result in a better performance.

The automatic translation of Deep Neural Network models into FPGA accelerators often relies on well-established High-Level Synthesis techniques to generate Verilog/VHDL code.
Whether they start from a pre-trained, unspecialized model or choose a hardware/network co-design approach, existing workflows are often composed of two steps.
First, the high-level Neural Network model is translated into a lower-level software programming language, and then High-Level Synthesis is applied to this intermediate representation to obtain an RTL design, ready to be synthesized and deployed on FPGA.

hls4ml \cite{duarte2018fast} follows this pattern: it is a toolchain designed to exploit High-Level Synthesis in the implementation flow for a Deep Neural Network accelerator on FPGA, and its intermediate format is C++ code.
Section \ref{sec:hls4ml} analyzes hls4ml more in detail since it is the baseline upon which our work is going to be developed.

FINN \cite{blott2018finn} is a compiler framework targeting Xilinx FPGAs that can produce high-throughput hardware accelerators and help explore architectural trade-offs.
Its primary focus is on quantized Deep Neural Networks: floating-point data are substituted with smaller, fixed-point types that are better suited to resource-limited FPGAs.
In particular, FINN works very well with extremely low bit-widths (less than 4 bits fixed-point), and it contains a Pytorch library for quantization-aware training to ensure reasonable accuracy even with such small data types.
Underneath the surface, FINN's approach is based on a library of parametric C components optimized for Vivado HLS.
The user can choose between two accelerator templates with different resource utilization and throughput characteristics: a feed-forward streaming dataflow pipeline (for small networks that can afford to store all weights on-chip) or a computation engine that offloads part of the data to external memory (for larger networks or smaller boards).
FINN is a powerful tool because it integrates each step of the design and implementation flow, from the definition and training of the network to the deployment of the accelerator on AWS FPGA instances.
Nevertheless, the compilation passes are designed for quantization-aware models with extremely low bit-widths, resulting in a network/hardware co-design approach that is not directly applicable to a generic pre-trained Deep Neural Network.

LeFlow \cite{noronha2018leflow} is another HLS-based project supporting the automatic implementation of Deep Learning operators on FPGA.
The input is a Tensorflow graph, transformed by Google's XLA compiler into an LLVM IR that can be processed by the LegUp HLS tool.
The obtained accelerators represent typical Neural Network layers, working under the assumption that both weights and inputs of the model are available on-chip: this strongly limits the possibility of using the same toolchain on complete networks.
Moreover, XLA is built to generate code for CPUs and GPUs, resulting in LLVM IRs that are not suitable for FPGAs and too low-level to apply meaningful optimizations.

Other tools exist that offer automatic methods to deploy Deep Neural Networks on FPGAs, as the ones described in \cite{venieris2018toolflows}.
Most of them support fixed-point data types arguing that the additional accuracy of floating-point calculation does not compensate for the resource utilization overhead; some are based on HLS, while others provide hand-written RTL kernels.
They are all focused on accelerating Convolutional Neural Networks, which are the backbone of computer vision tasks, so they benefit from the regular computation patterns, redundancy, and reduced dynamic range typical of convolutional layers.

\section{hls4ml: benefits and drawbacks} \label{sec:hls4ml}

hls4ml was initially presented as a compiler for the implementation of fully connected Neural Networks on FPGA \cite{duarte2018fast}, presenting a specific use case where a Multi-Layer Perceptron was used to perform a classification task within a physics experiment at CERN.
hls4ml automatically translates high-level models built and trained within typical Deep Learning frameworks (Keras, Tensorflow, Pytorch, or any other that can export graphs in the ONNX format) into C++ code and then uses Vivado HLS to generate a corresponding RTL design.
The translation to C++ is based on a library of parametric templates that work with fixed-point data types, effectively quantizing the original model post-training.
Other optimizations introduced to comply with the strict latency constraints imposed by the experiment are weights compression (pruning) and parallelization of multiplications.
Following studies presented additions to the tool to support quantization-aware training and automatic heterogeneous quantization \cite{coelho2020ultra}, binary and ternary networks \cite{ngadiuba2020bintern}, and Convolutional Neural Networks \cite{aarrestad2021zos}.

A user-friendly interface based on Python makes hls4ml accessible to Deep Learning developers that are not familiar with hardware design.
Key optimization choices are exposed as configuration options: the user can specify a data type for the whole model or on a per-layer basis and tune parallelism against resource usage for multipliers (reuse factor).
Network compression is exploited by enforcing sparsity in the training phase and relying on Vivado HLS to eliminate operations with zero-weights, so there is no configuration option to control pruning.
After the user imported a model and specified configuration options, hls4ml parses the model and builds a C++ representation: it selects the corresponding operators from a library of parametric components, adjusts data types for weights, biases, and results, and adds compilation directives (pragmas) to drive the High-Level Synthesis process according to the user-defined configuration.
Finally, hls4ml takes care of launching Vivado HLS and logic synthesis tools to generate the design and evaluate its performance.

Alongside the evident benefits that hls4ml can offer in the automation of what would otherwise be a long, expensive manual translation, there are some drawbacks that inherently limit its usage.
Appropriate implementation choices for small, fully-connected models under tight latency constraints may not always be beneficial for larger networks: most notably, storing network weights inside on-chip logic and unrolling all loops to increase parallelism quickly depletes available resources.
This approach does not scale well to state-of-the-art Deep Neural Networks, having orders of magnitude more weights and computations than the 3-layer MLP model presented in the original hls4ml publication.
Moreover, optimizing the C++ library for Vivado HLS prevents the use of hls4ml to develop accelerators on FPGAs from vendors other than Xilinx.
To the best of our knowledge, support to target Intel FPGAs is under development, but it may require a new implementation of the components library, reinforcing our hypothesis that portability is a weak point for hls4ml.

The library implementation of activation functions contains a practical example of features that are only applicable on a Xilinx backend through Vivado HLS.
Non-trivial activation functions (including softmax) are implemented as tables of constant values representing the function output for a given range of inputs.
To create the tables, the C++ library initializes a list of values following a pattern that Vivado HLS recognizes and transforms at compile-time so that the final RTL design only contains the pre-computed lookup table stored in BRAMs.
Instead, other HLS tools may not interpret the C++ pattern as the construction of a constant table, wasting resources and increasing the overall latency to implement the actual mathematical functions and control logic contained in the code.

While investigating this issue, we also found out that if the configuration defines a default fixed-point type for the entire network, the decision is overwritten in the softmax layer with a fixed-point type empirically selected to fill the available BRAM size on specific Xilinx FPGA models.
In particular, to target a Xilinx 18k BRAM, tables of 1024 elements encoded as 18-bit fixed-point values are used.

\section{Proposed improvements} \label{sec:improvements}

Our work aims at overcoming some limitations of the hls4ml framework to increase its efficiency and applicability: the front-end to import models will mostly remain unchanged to maintain a user-friendly interface, while the library of C++ components will need significant modifications.
In this first phase, we conducted an in-depth analysis of hls4ml and identified key opportunities for improvement: some of those are still preliminary ideas, and they will need further study and refinement; for others, we defined a path towards implementation.
The following sections describe the two areas where we propose to focus our efforts.

\subsection{Portability}

hls4ml uses Vivado HLS as back-end to generate RTL designs from C++ code, and this choice strongly influences the library of components used to build the C++ representation of the input Neural Network model.
Each component is described with coding techniques and compiler directives (pragmas) specific to Vivado HLS, limiting the portability of code to other High-Level Synthesis tools.
Efforts to make the library less specific would enable using different HLS and logic synthesis tools in the back-end, extending the range of supported FPGA platforms: our goal is to leverage an open-source HLS tool, PandA-bambu \cite{panda}, but the de-specialization of the library would also open the way for other commercial and open-source HLS tools.

Modifications to the library have already started: as described in Section \ref{sec:hls4ml}, the implementation of activation functions as tables of constant values depended on a special feature of Vivado HLS, so we had to find a different solution to replicate the same behavior in a way that other tools could correctly interpret.
Arrays of constant values can be computed exploiting the concept of `constexpr,' introduced in the C++11 standard \cite{constexpr} to identify functions and variables that can be evaluated at compile-time, together with some functions of the C++14 standard library.
In our new implementation, each activation function contains a static method called `compute()' with instructions for the computation of the desired function output.
A dedicated class template is called to initialize the constant tables, requiring as input the `compute()' method and the length N of the array that will be constructed: at compile-time, a constexpr array of N elements will be created and populated with values calculated by the specific `compute()' method.
This version of the code correctly produces tables of constant values thanks to compile-time expression resolution, and it does not need to rely on Vivado HLS anymore.

The realization of this solution highlighted some additional issues: for example, hls4ml uses data types from the arbitrary precision library by Xilinx (ap\_types \cite{aptypes}), which cannot be used within a constexpr.
Instead of modifying the ap\_types library, we preferred to replace it with our extension of a similar open source library based on the "Algorithmic C (AC)" data types from Mentor Graphics \cite{actypes}.
Both libraries are designed for functional simulation, and they have separate synthesis-oriented implementations that are only compatible with a specific High-Level Synthesis tool: our modified version of ac\_types instead is already optimized for synthesis, and it can be compiled with standard C/C++ compilers.
With a few modifications, the types defined by our library can be used within constexpr.
Finally, the original implementations of the activation functions were based on mathematical functions available in the C++ standard library (exponential, trigonometric, and hyperbolic functions), which were also not compatible with constexpr, so it was necessary to substitute them with the constexpr-compliant math library `gcem' \cite{gcem}.

\subsection{Quantization}

Together with the low power consumption, speed, and flexibility, another feature offered by FPGAs is the possibility of exploring custom data types, exposing opportunities for quantization that are not available or inefficient on general-purpose processors.
Neural Networks are inherently robust to errors, so quantization to integer and fixed-point data types has been widely studied and applied, with different techniques and on different hardware platforms.
The underlying assumption is that a small percentage loss in accuracy is well compensated by significant reductions in power consumption, memory requirements, and area utilization: floating-point data types are considered essential during the training phase but inefficient for inference.

However, it is reasonable to assume that not all application scenarios that would benefit from FPGA acceleration are tolerant to the accuracy loss introduced by integer and fixed-point quantization (especially if applied post-training).
Custom floating-point data types, based on the IEEE standard or with entirely different numeric encodings, can disclose a new design space where the accuracy is improved with respect to fixed-point quantization and performances are better than baseline floating-point models.

In hls4ml, fixed-point quantization is the default choice: the user can select the most appropriate fixed-point formats for each layer of the input model, evaluating through simulation what is their impact on accuracy and performance.
The underlying assumption is that FPGA implementations of floating-point operations are usually inefficient (Vivado HLS itself often discourages the use of floating-point data types).
Instead, we want to extend the library of components to allow the use of customized floating-point data types in hls4ml: we have full control over the internal behavior of bambu, so it will be possible to match custom data types in the C++ representation with optimized RTL operators.
The front-end to import models will not require substantial modifications, while support for functional simulation will be added to measure accuracy loss; in the future, it could also be useful to introduce automatic design space exploration tools to direct the choice of data types.

\section{Conclusion}

hls4ml is a useful tool to simplify the translation of Deep Neural Network models into FPGA accelerators: its current version can produce optimized designs with low latency for a small class of input models, targeting Xilinx FPGAs through Vivado HLS.
In this paper, we outlined a plan to modify the C++ library at the core of hls4ml to remove some of its limitations: we expect that our de-specialization effort will increase the applicability to a more diverse set of input applications and portability to a broader selection of FPGA targets, while at the same time leading to the creation of more efficient accelerators.

\end{document}